\documentclass[letterpaper,journal]{IEEEtran}
\usepackage{amsmath,amsfonts}
\usepackage{algpseudocode}
\usepackage{algorithm}
\usepackage{array}
\usepackage{graphicx}
\usepackage{amsmath,amssymb,amsfonts}
\usepackage[caption=false,font=normalsize,labelfont=sf,textfont=sf]{subfig}
\usepackage{textcomp}
\usepackage{stfloats}
\usepackage{url}
\usepackage{verbatim}
\usepackage{graphicx}
\usepackage{graphicx}
\usepackage{subcaption}
\usepackage[caption=false]{subfig}
\usepackage{cite}
\usepackage{newtxtext,newtxmath} 
\usepackage[caption=false,font=normalsize]{subfig}
\usepackage{multirow}
\usepackage{acronym}
\newacro{OFDM}{Orthogonal Frequency Division Multiplexing}
\newacro{DLI}{Direct Link Interference}
\newacro{DL}{Direct-Link}
\newacro{ELPC}{Extremely Low-Power Communications}
\newacro{BER}{Bit Error Rate}
\newacro{mMTC}{Massive Machine Type Communications}
\newacro{IoT}{Internet of Things}
\newacro{RF}{Radio Frequency}
\newacro{SINR}{Signal to Interference Plus Noise Ratio}
\newacro{CP}{Cyclic Prefix}
\newacro{SIC}{Successive Interference Cancellation}
\newacro{CSI}{Channel State Information}
\newacro{BR}{Backscatter Receiver}
\newacro{BD}{Backscatter Device}
\newacro{SAW}{Surface Acoustic Wave}
\newacro{3GPP}{Third Generation Partnership Project}
\newacro{BL}{Backscatter-Link}
\newacro{OTSM}{Orthogonal Time Sequency Multiplexing}
\newacro{AWGN}{Additive White Gaussian Noise}
\newacro{CIR}{Channel Impulse Response}
\newacro{FIR}{Finite Impulse Response}
\newacro{BPSK}{Binary Phase Shift Keying}
\newacro{DFT}{Discrete Fourier Transform}
\newacro{SNR}{Signal-to-Noise Ratio}
\newacro{AmBC}{Ambient Backscatter Communication }
\newacro{BDs}{Backscatter Devices}
\newacro{A-IoT}{Ambient Internet of Things}
\newacro{BPSK}{Binary Phase Shift Keying}
\newacro{CLT}{Central Limit Theorem}
\newacro{IDFT}{Inverse Discrete Fourier Transform}
\newacro{FR1}{Frequency Range 1}
\newacro{S}{Ambien OFDM Source}
\newacro{R}{Receiver}
\usepackage{xcolor}
\usepackage{soul}
\usepackage{amsmath}
\usepackage{amssymb}
\PassOptionsToPackage{colorlinks=true,linkcolor=blue,citecolor=blue,urlcolor=blue}{hyperref}
\usepackage{orcidlink}

\def\BibTeX{{\rm B\kern-.05em{\sc i\kern-.025em b}\kern-.08em
T\kern-.1667em\lower.7ex\hbox{E}\kern-.125emX}}

\begin{document}
\title{Shuf-Scatter: Permutation-Based DLI Suppression for OFDM Ambient Backscatter}
\author{Salim Chabani, Riadh Ouzane, Salah Eddine Zegrar,~\IEEEmembership{Member,~IEEE},
and H\"{u}seyin Arslan,~\IEEEmembership{Fellow,~IEEE}
\thanks{Salim Chabani, and H\"{u}seyin Arslan are with the Department of Electrical and Electronics Engineering,
Istanbul Medipol University, Istanbul 34810, Turkey. E-mail: salim.chabani@std.medipol.edu.tr;
huseyinarslan@medipol.edu.tr.}%
\thanks{Riadh Ouzane is with the Department of Electrical Engineering and Computer Engineering, Laval University, Québec, QC, Canada. E-mail: riadh.ouzane.1@ulaval.ca.}%
\thanks{Salah Eddine Zegrar is with the Computer, Electrical and Mathematical Sciences and Engineering (CEMSE) Division, KAUST University, Thuwal, Saudi Arabia. E-mail:
salaheddine.zegrar@kaust.edu.sa}%
}

\maketitle

\begin{abstract}
Direct Link Interference  (DLI) remains a major bottleneck in Orthogonal Frequency Division Multiplexing (OFDM)-based Ambient Backscatter Communication (AmBC) systems. This letter proposes Shuf-Scatter, a scheme that applies a deterministic permutation to the reflected \ac{OFDM} waveform to create a distinct correlation-domain signature. By exploiting the quasi-orthogonality between the reshuffled backscatter signal and the direct-link component, the proposed receiver effectively mitigates \ac{DLI} and enables reliable detection without requiring dedicated spectral resources, or instantaneous  Backscatter-Link (BL) Channel State Information (CSI). Analytical and simulation results demonstrate that Shuf-Scatter achieves superior  Bit Error Rate (BER) performance under strong \ac{DLI} while preserving the integrity of the legacy \ac{OFDM} transmission.
\end{abstract}

\section{Introduction}
\ac{AmBC} enables ultra-low-power and low-cost devices to communicate by modulating and reflecting existing radio-frequency signals instead of generating dedicated carrier waveforms. This capability makes \ac{AmBC} a promising solution for battery-less
and energy-constrained \ac{BD}~\cite{8368232}. However, the weak \ac{BL} signal is received alongside the much stronger \ac{DL} transmission from the ambient \ac{OFDM} source. The resulting \ac{DLI} often dominates the received signal, making reliable detection of the backscattered information challenging.

Several interference handling approaches have been proposed for \ac{AmBC} systems. In~\cite{devineni2021noncoherent}, averaging-based schemes exploit the lower symbol rate of the \ac{BD} to suppress the fast-varying direct-link component. The authors in~\cite{yang2017modulation,s19030517} exploit the intersymbol interference-free region of the \ac{CP} in \ac{OFDM} systems as a detection window for backscatter signal recovery. However, the limited \ac{CP} length restricts the available detection window; consequently, the achievable performance. Pilot-aided and coherent schemes estimate the cascaded backscatter channel using known \ac{OFDM} resources~\cite{kishore2021pilot}, whereas Noncoherent and \ac{SIC}-based methods have also been investigated\cite{Qian2016Noncoherent,10236556}. However, due to the large power disparity between the direct-link and backscatter signals, even small channel estimation, decoding, or cancellation errors can generate residual interference comparable to the desired backscatter signal, thereby degrading detection performance, particularly in low-\ac{SNR} regimes.

Alternative designs, including null-subcarrier and frequency-shift-based modulation, exploit orthogonal frequency-domain resources to separate the backscatter signal from \ac{DLI}-dominated carriers~\cite{tao2019mfsk,nullsubcarrier}. Waveform-level approaches, such as \ac{OTSM}-based backscatter modulation, have also been proposed to enhance robustness against interference and doubly selective channels~\cite{Ouzane2025OTSM}. Nevertheless, \ac{OFDM} remains the dominant waveform in next-generation wireless systems and is actively supported by \ac{3GPP} and other standardization bodies for \ac{BD} deployments~\cite{3GPP2024AmbientIoT}. Consequently, developing \ac{OFDM}-compatible backscatter solutions remains highly relevant for practical \ac{AmBC} deployments. More recently, learning-assisted and beamforming-based schemes have demonstrated enhanced \ac{DLI} suppression capabilities~\cite{10673898,10000927}. However, these approaches typically rely on multi-stage training procedures and large antenna arrays, limiting their suitability for low-power \ac{A-IoT} applications.

Motivated by the limitations of existing \ac{DLI} mitigation approaches~\cite{devineni2021noncoherent,yang2017modulation,s19030517,kishore2021pilot,Qian2016Noncoherent,10236556,tao2019mfsk,nullsubcarrier,Ouzane2025OTSM,3GPP2024AmbientIoT,10000927,10673898}, this letter proposes Shuf-Scatter, a permutation-based backscatter mechanism that deterministically reorders adjacent segments of the incident \ac{OFDM} waveform at the \ac{BD}. The corresponding receiver applies the same reshuffling to the recovered direct-link waveform and performs normalized correlation in the time domain to distinguish the backscatter component from the dominant \ac{DLI}. As a result, Shuf-Scatter enables backscatter link detection while preserving the legacy \ac{OFDM} transmission.
\footnote{\textbf{Notation:}
Non-bold uppercase and lowercase letters denote scalar quantities.$\lfloor\cdot\rfloor$ denotes the floor operation, and $\mathbb{E}\{\cdot\}$ denotes expectation. A circularly symmetric complex Gaussian random variable is denoted by $\mathcal{CN}(0,\sigma^2)$. The Dirac delta function is $\delta[\cdot]$. The modulo operation is defined as $[\ell]_{\beta}= \ell \bmod \beta$. The operator $\ast$ and $\circledast$ denote linear/circular convolution, respectively.}

\section{System Model}
\begin{figure}[!t]
\centering
\includegraphics[width=1\columnwidth]{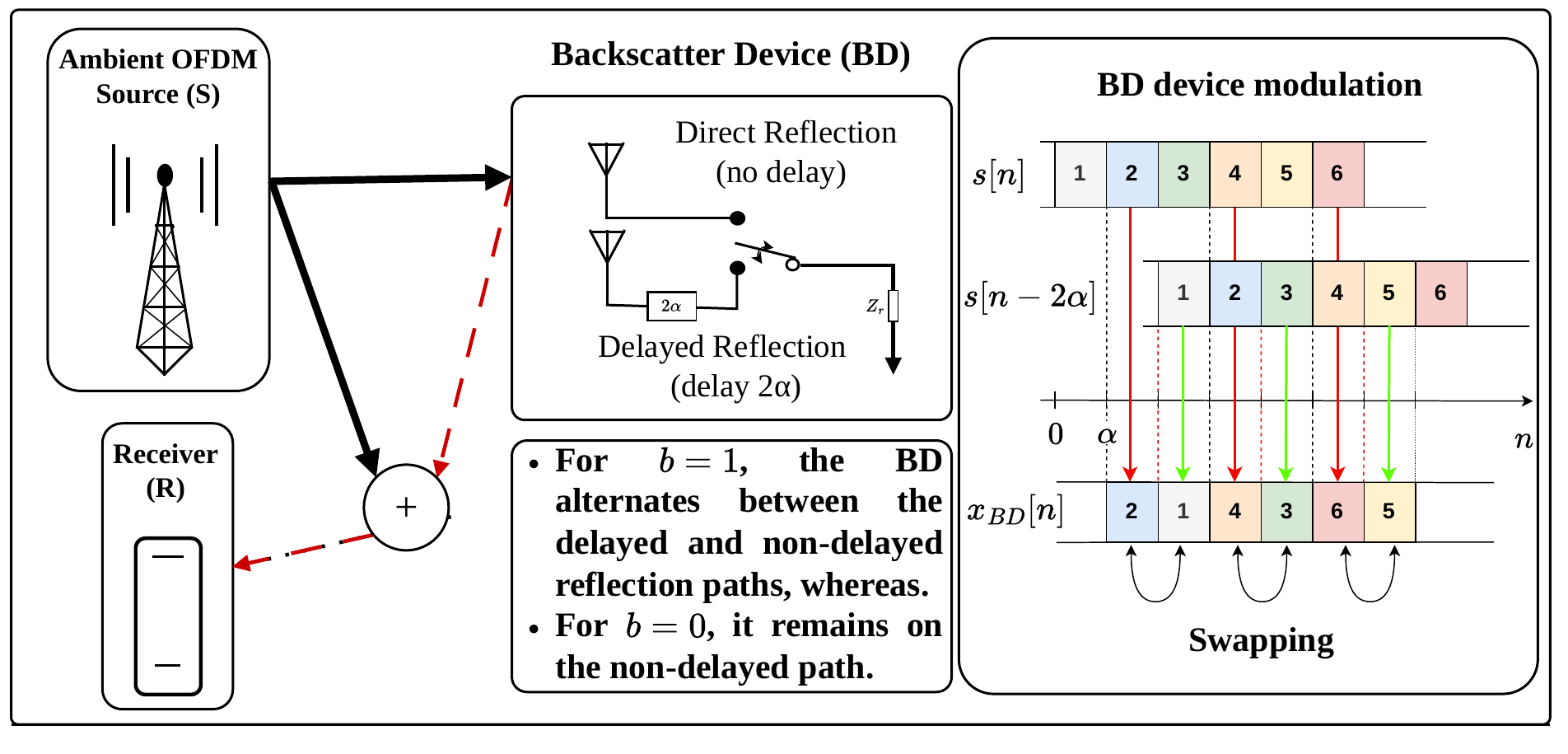}
\caption{System model of the proposed Shuf-Scatter AmBC system.}
\label{fig:bd_circuit}
\end{figure}

We consider the Shuf-Scatter \ac{AmBC} system shown in Fig.~\ref{fig:bd_circuit}, where an ambient \ac{S} communicates with the \ac{R} while simultaneously illuminating a passive \ac{BD}, thereby creating a direct-link signal at the receiver. The source-to-receiver (\ac{S}-\ac{R}) channel is denoted by $h_{\mathrm{d}}$, whereas the equivalent cascaded source-\ac{BD}-receiver (\ac{S}-\ac{BD}-\ac{R}) channel is denoted by $h_b = h_{\mathrm{BR}} * h_{\mathrm{SB}}$.

\subsection{Ambient OFDM Signal Model}
A conventional \ac{CP}-\ac{OFDM} waveform is transmitted by the source \ac{S}, which serves as the excitation signal for the passive \ac{BD}. Considering an \ac{OFDM} symbol with $N$ subcarriers, the corresponding time-domain signal is obtained through an $N$-point \ac{IDFT} as
\begin{equation}
x[\hat{n}] =
\frac{1}{\sqrt{N}}
\sum_{k=0}^{N-1}
X[k]\exp\left(j\frac{2\pi k\hat{n}}{N}\right),
\qquad
\hat{n} = 0,\ldots,N-1,
\label{eq:ofdm_time_domain}
\end{equation}
where $X[k]$ denotes the data symbol transmitted on the $k$-th subcarrier. Finally, a \ac{CP} of length $N_{\mathrm{cp}}$, chosen to exceed the maximum channel delay spread $\ell_{b,\max}$, is appended prior to transmission.
\subsection{Propagation Channel Model}
The wireless propagation environment is modeled using quasi static \ac{FIR} channels. Accordingly, the discrete-time impulse response of link $\eta\in\{\mathrm{d},\mathrm{b}\}$ is given by
\begin{equation}
h_{\eta}[n]
=\sum_{i=0}^{L_{\eta}-1}
g_{\eta,i}\delta[n-\ell_{\eta,i}],
\label{eq:fir_channel_model}
\end{equation}
where $\ell_{\eta,i}$ denotes the delay of the $i$th propagation path and $g_{\eta,i}$ its corresponding complex gain. The distribution of $g_{\eta,i}$ depends on the considered link. For the direct S-R channel, $g_{\mathrm{d},i}\sim\mathcal{CN}(0,\sigma_{\mathrm{d}}^{2})$, whereas for the cascaded S-\ac{BD}-R channel, the double-fading effect is modeled by $g_{\mathrm{b},i}\sim\mathcal{DCN}(0,\sigma_{\mathrm{b}}^{2})$~\cite{PLS_3gpp}. \footnote{The considered TDL-A channel model follows the link-level evaluation assumptions specified in \cite[Table 4.2.2-1]{3GPP2024AmbientIoT} for \ac{FR1} deployments and represents a low-mobility (3 km/h), short-range, rich-scattering indoor-factory propagation environment.}

\subsection{Proposed \ac{BD} Modulation Scheme}
\label{subsec:bd_modulation_model}
Unlike conventional symbol-level backscatter modulation, the proposed scheme conveys information by altering the temporal structure of the reflected \ac{OFDM} waveform over intervals of length $\alpha$.
For $b=0$, the incident \ac{OFDM} symbol is reflected without alteration,i.e.,
\begin{equation}
    x_{\mathrm{BD}}^{(0)}[n] = s[n], \quad \text{ for }~ n= 0,\cdots,(N+N_{\mathrm{cp}})-1.
    \label{eq0}
\end{equation}
whereas, for $b=1$, as shown in Fig.~\ref{fig:bd_circuit} the \ac{BD} alternates every $\alpha$ samples between the direct and delayed reflection paths, this operation exchanges each pair of adjacent $\alpha$-sample segments, so that the second segment of the pair is reflected before the first one. As illustrated by the example on the right side of Fig.~\ref{fig:bd_circuit}. Hence, the incident
\ac{OFDM} waveform is deterministically reshuffled, yielding
\begin{equation}
    x_{\mathrm{BD}}^{(1)}[n]=
\begin{cases}
s\left(\big[n+\alpha\big]_{N+N_{\mathrm{CP}}}\right),
& 2m\alpha \le n < (2m+1)\alpha, \\
s\left(\big[n-\alpha\big]_{N+N_{\mathrm{CP}}}\right),
& (2m+1)\alpha \le n < 2(m+1)\alpha,
\end{cases}
\label{eq:1}
\end{equation}
for $$m=0,\ldots,\frac{N+N_{\mathrm{CP}}}{2\alpha}-1,$$
where $\alpha\in\mathbb{Z}^{+}$ represents the switching interval in samples and sets the length of each exchanged waveform segment. For analytical tractability, we assume that
$\ell_{b,\max}+\alpha<N_{\mathrm{CP}}$ and that $N+N_{\mathrm{CP}}$ is an integer multiple of $2\alpha$, so that the \ac{OFDM} block is partitioned into complete pairs of $\alpha$-sample intervals. 

\textbf{\textit{Remark 1:}} The apparent $+\alpha$ shift in \eqref{eq:1} is an indexing representation of the adjacent-segment exchange and does not imply access to future samples. As illustrated in Fig. \ref{fig:bd_circuit}, the 
causal implementation uses direct and delayed branches separated by a relative delay of $2\alpha$ samples. Their alternating selection produces an effective delay of only $\alpha$ samples.

\begin{figure}[!t]
\centering
\includegraphics[
    width=1\columnwidth,
    trim=1.5cm 0.7cm 1.9cm 1cm,
    clip
]{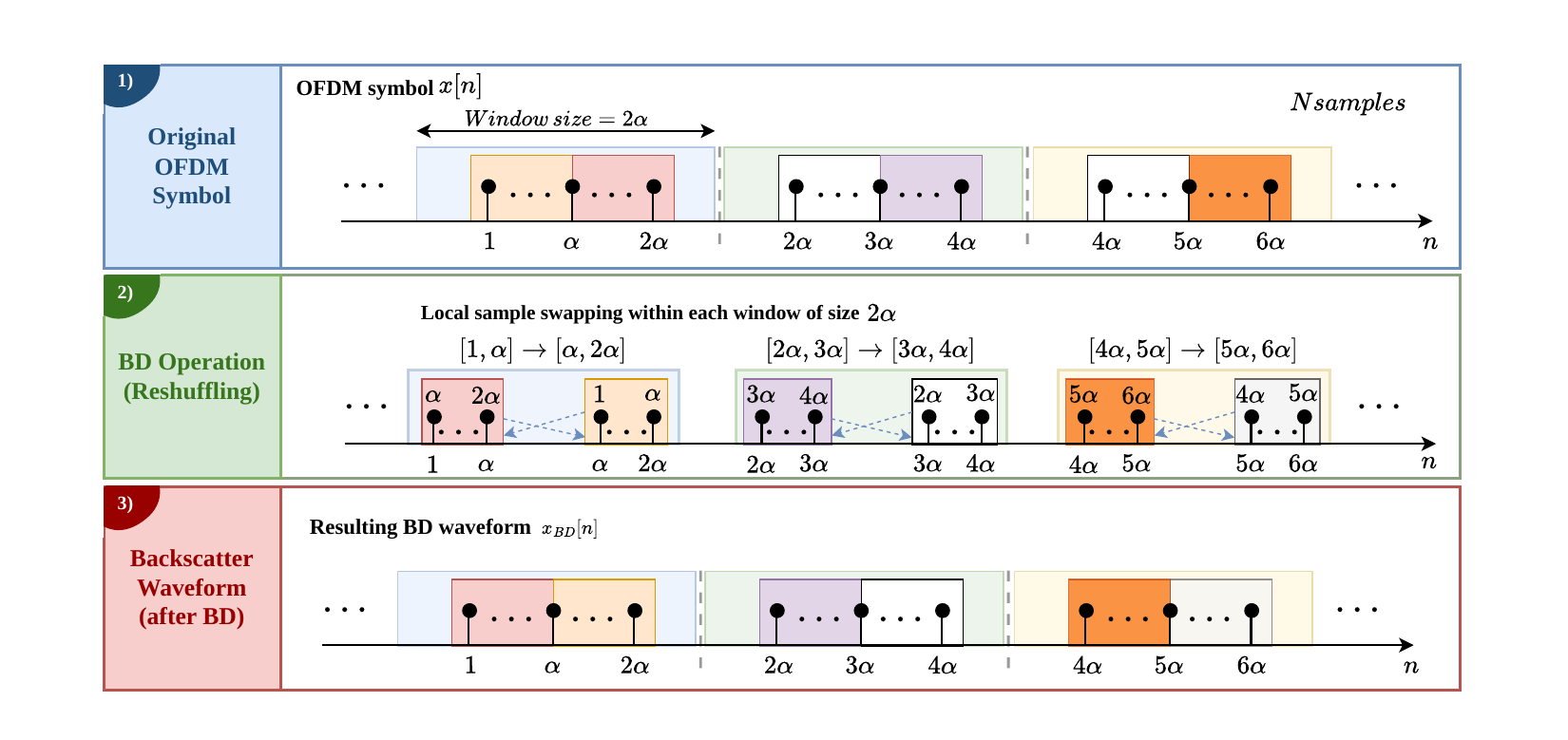}
\caption{Proposed Shuf-Scatter \ac{BD} modulation.}
\label{fig:illustration}
\end{figure}

By combining the modulation states defined in \eqref{eq0} and \eqref{eq:1}, the proposed waveform-control operation can be written in a compact form as
\begin{equation}
\begin{aligned}
x_{\mathrm{BD}}^{(b)}[n]=&(1-b)s[n] +bs\Big[\big[ n+(-1)^{\left\lfloor n/\alpha \right\rfloor}\alpha \big]_{(N+N_{\mathrm{CP}})} \Big],
\\& \text{for} \quad n=0,1,\ldots,(N+N_{\mathrm{CP}})-1.
\end{aligned}
\label{eq:bd_modulation}
\end{equation}
Here, $\lfloor n/\alpha \rfloor$ denotes the index of the $\alpha$-sample interval containing sample $n$; hence, $(-1)^{\lfloor n/\alpha \rfloor}$ alternates between $+1$ and $-1$ every $\alpha$ samples. An illustrative example is shown in Fig.~\ref{fig:illustration}.

\textbf{\textit{Remark 2:}} For the modulation state $b=1$, the proposed \ac{BD} alternates between its two reflection states every $\alpha$ samples. Consequently, the effective switching rate is given by
\begin{equation}
f_{\mathrm{BD}}= \frac{f_s}{\alpha},
\label{eq:switching_rate}
\end{equation}
where $f_s$ denotes the sampling frequency of the ambient \ac{OFDM} waveform. Therefore, increasing $\alpha$ reduces the switching rate and the number of state transitions within an \ac{OFDM} symbol. As a result, the temporal variation introduced by the \ac{BD} becomes less pronounced, reducing the distinguishability between the \ac{DL} and \ac{BL} components at the receiver.

\subsection{On the Quasi-Orthogonality of Shuf-Scatter Modulation}
\label{sec:quasi_orthogonality}
The quasi-orthogonal behavior between the original and reshuffled \ac{OFDM} waveforms fundamentally stems from the statistical randomness of the independent subcarriers. In this subsection, we characterize the Shuf-Scatter correlation structure and its deterministic lag positions. Accordingly, the analysis considers the case $b=1$ while neglecting the impact of the wireless channel for simplicity. Let $r$ denote an arbitrary time lag where $r\in\{0,1,\ldots,N-1\}$, and define the instantaneous OFDM-block energy as $E_X=\sum_{\ell=0}^{N-1}|X[\ell]|^2$. The normalized cross-correlation between the original and reshuffled waveforms is defined as
\begin{equation}\small
\begin{aligned}
\rho[r; x,x_{\mathrm{BD}}] =& \frac{\sum_{n=0}^{N-1}x^*[n]x_{\mathrm{BD}}[n-r]}{\sqrt{\sum_{n=0}^{N-1}|x[n]|^2}\cdot \sqrt{\sum_{n=0}^{N-1}|x_{\mathrm{BD}}[n-r]|^2}}\\
=&\frac{1}{E_X}\sum_{n=0}^{N-1}x^{*}[n]x_{\mathrm{BD}}[n-r].
\end{aligned}
\label{eq:normalized_correlation}
\end{equation}
By substituting the \ac{OFDM} signal model in \eqref{eq:ofdm_time_domain} and the reshuffled waveform definition in \eqref{eq:bd_modulation} into \eqref{eq:normalized_correlation}, we obtain

\begin{equation}
\begin{aligned}
\rho[r;x,x_{\mathrm{BD}}]
&=\frac{1}{NE_X}\sum_{n=0}^{N-1}\sum_{k=0}^{N-1}
\sum_{m=0}^{N-1}X^{*}[k]X[m]\\
&\quad\times\exp\left\{j\frac{2\pi}{N}
\Big[(m-k)n-mr+m\alpha(-1)^{\left\lfloor \frac{n-r}{\alpha} \right\rfloor}
\Big]
\right\}.
\end{aligned}
\label{eq:rho_ofdm_expanded}
\end{equation}

For independent, zero-mean symbols drawn from a conventional symmetric PSK/QAM constellation,
$\mathbb{E}\{\frac{X^*[k]X[m]}{E_X}\}=0$ for $k\neq m$, while symmetry across the identically distributed subcarriers gives $\mathbb{E}\{\frac{|X[k]|^2}{E_X}\}=\frac{1}{N}$. Hence, taking the expectation of \eqref{eq:rho_ofdm_expanded} eliminates the $k\neq m$ terms and, with $m=k$, yields
\begin{equation}
\mathbb{E}\left\{\rho[r;x,x_{\mathrm{BD}}]\right\}
=\frac{1}{N^2}\sum_{n=0}^{N-1}\sum_{k=0}^{N-1}
e^{j\frac{2\pi k}{N}q_n(r,\alpha)},
\label{double_sum}
\end{equation}
where $q_n(r,\alpha)=-r+\alpha(-1)^{\left\lfloor \frac{n-r}{\alpha}\right\rfloor}$.
Clearly, from \eqref{double_sum}, the inner summation is a finite Fourier sum \cite{5467212}; hence, we have
\begin{equation}
\sum_{k=0}^{N-1}
e^{j\frac{2\pi k}{N}q_n(r,\alpha)}
=\begin{cases}
N, & q_n(r,\alpha)=[0]_N,\\
0, & \mathrm{otherwise}.
\end{cases}
\end{equation}
Then, \eqref{double_sum} can be reformulated as
\begin{equation}
\mathbb{E}\left\{\rho[r;x,x_{\mathrm{BD}}]\right\}
=\frac{1}{N}\sum_{n=0}^{N-1}
\delta\left[-r+\alpha(-1)^{\left\lfloor \frac{n-r}{\alpha}\right\rfloor}\right]_N.
\label{final}
\end{equation}
It can be seen from \eqref{final} that a nonzero mean correlation occurs only when $-r+\alpha(-1)^{\left\lfloor \frac{n-r}{\alpha}\right\rfloor}=[0]_N$. Since $(-1)^{\left\lfloor (n-r)/\alpha\right\rfloor}\in\{+1,-1\}$, the constructive lags are limited to $r=\alpha$ and $r=N-\alpha$. Hence, the mean correlation is confined to these deterministic lags, which bound the interval within which the \ac{BD} signature is exploited for subsequent \ac{BL} signal detection.

\subsection{Received Signal}
At the receiver side, after removing the \ac{CP} the received time-domain \ac{OFDM} symbol can be written as
\begin{equation}\small
\begin{aligned}
y[n]=& x[n]\circledast h_d[n]+\beta x_{\mathrm{BD}}^{(b)}[n]\circledast h_b[n]+w[n] \\
=& \sum_{i=0}^{L_d-1} g_{d,i} x[n\ell_{d,i}]+\beta\sum_{i=0}^{L_b-1} g_{b,i}x_{\mathrm{BD}}^{(b)}[n-\ell_{b,i}]+w[n],\\=& \sum_{i=0}^{L_d-1} g_{d,i}x[n-\ell_{d,i}]+\beta\Big((1-b)\sum_{i=0}^{L_b-1} g_{b,i}x[n-\ell_{b,i}] \\
&+b\sum_{i=0}^{L_b-1} g_{b,i}x\left[n-\ell_{b,i}+(-1)^{\left\lfloor \frac{n-\ell_{b,i}}{\alpha}\right\rfloor}\alpha\right]_{N}\Big)+w[n],
\end{aligned}
\label{eq:rx_signal}
\end{equation}
where $y[n]\in\mathbb{C}^{\mathrm{N}}$, $\beta$ is the reflection coefficient of the \ac{BD} and $w\sim\mathcal{CN}(0,\sigma_w^2)$ is the \ac{AWGN}.

\section{SHUF-SCATTER SIGNAL RECOVERY }
This section presents the receiver processing for recovering the backscatter information by exploiting the quasi-orthogonality property introduced by the proposed Shuf-Scatter modulation.
\subsection{Backscatter Signal Detection}
Given that the backscattered signal power is orders of magnitude below that of the direct link, the receiver can recover the dominant direct-link \ac{OFDM} waveform $\hat{x}[n]$ considering the weak \ac{BL} component as part of the \ac{AWGN}\cite{Ouzane2025OTSM}, given as
 \begin{equation}
 \hat{y}[n]=\sum_{i=0}^{L_d-1} g_{d,i}x[n-\ell_{d,i}]+\tilde{w}[n],
 \label{eq:before}
 \end{equation}
 where $\tilde{w}[n]=\beta\sum_{i=0}^{L_b-1} g_{b,i}x_{\mathrm{BD}}^{(b)}[n-\ell_{b,i}]+w[n]$.
 Applying the $N$-point \ac{DFT} gives the frequency-domain model as
\begin{equation}\small
\begin{aligned}
\hat{Y}[k] &=\sum_{n=0}^{N-1} \sum_{i=0}^{L_{\mathrm{d}}-1} g_{d,i} x[n-\ell_{d,i}] e^{-j\frac{2\pi k}{N}n} +\widetilde{W}[k] \\&=\sum_{i=0}^{L_{\mathrm{d}}-1} g_{d,i} e^{-j\frac{2\pi k\ell_{d,i}}{N}} \sum_{n=0}^{N-1} x[n] e^{-j\frac{2\pi kn}{N}} + \widetilde{W}[k]\\&=H_{\mathrm d}[k]X[k]+\widetilde{W}[k],
\label{eq:FD}
\end{aligned}
\end{equation}
where $H_d[k], X[k]$ and $\tilde{W}[k]$ are the frequency domain representation of $h_d[n], x[n]$ and $\tilde{w}[n]$, respectively. The dominant \ac{OFDM} waveform is then equalized using single-tap equalizer followed by the \ac{IDFT} as
\begin{equation}
\begin{aligned}
\widehat{x}[n]
&=\frac{1}{\sqrt{N}}
\sum_{k=0}^{N-1}X[k]e^{j\frac{2\pi kn}{N}}+\frac{1}{\sqrt{N}}\sum_{k=0}^{N-1}
\frac{H_{\mathrm d}^{*}[k]}{|H_{\mathrm d}[k]|^{2}}
\widetilde{W}[k]e^{j\frac{2\pi kn}{N}}  \\
&=x[n]+\widetilde{v}[n],
\end{aligned}
\label{eq:xhat_time_noise}
\end{equation}
where $\widetilde{v}[n]=
\frac{1}{\sqrt{N}}\sum_{k=0}^{N-1}
\frac{H_{\mathrm d}^{*}[k]}{|H_{\mathrm d}[k]|^{2}}
\widetilde{W}[k]e^{j2\pi kn/N}$ denotes the residual time-domain error.
\begin{figure}[t!]
\centering
\resizebox{0.9\columnwidth}{!}{
\includegraphics{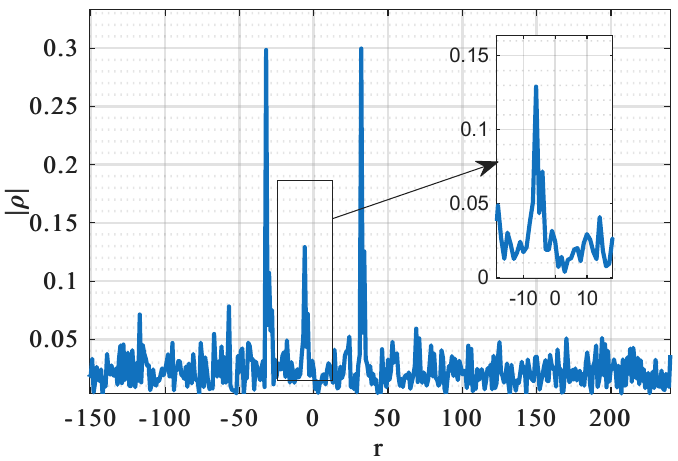}}
\caption{Normalized correlation between the received signal  and the locally generated shuffled reference for the bit-$1$ case.}
\label{fig:corr1}
\end{figure}
The recovered signal in \eqref{eq:xhat_time_noise} is reshuffled according to the same $\alpha$-sample permutation pattern employed by the \ac{BD}, yielding the locally generated reference signal $\hat{x}_{\mathrm{BD}}[n]$ given by
\begin{equation}
\hat{x}_{\mathrm{BD}}[n] = \hat{x}\left( \big[ n+(-1)^{\lfloor n/\alpha \rfloor}\alpha \big]_{N} \right).
\label{eq:shuf_scatter_ref}
\end{equation}
To extract the backscatter data, the receiver correlates the composite received signal $y[n]$ with $\hat{x}_{\mathrm{BD}}[n]$, yielding
\begin{equation}
\begin{aligned}
\rho[r;y,\hat{x}_{\mathrm{BD}}]
&= \sum_{i=0}^{L_{\mathrm{d}}-1} g_{\mathrm{d},i}^{*} \frac{1}{N} \sum_{n=0}^{N-1} x^{*}[n-\ell_{\mathrm{d},i}] \hat{x}_{\mathrm{BD}}[n-r] \\
&\quad + \beta \sum_{i=0}^{L_{\mathrm{b}}-1} g_{\mathrm{b},i}^{*} \frac{1}{N} \sum_{n=0}^{N-1} \left(x_{\mathrm{BD}}^{(b)}[n-\ell_{\mathrm{b},i}]\right)^{*} \hat{x}_{\mathrm{BD}}[n-r] \\
&\quad + \frac{1}{N} \sum_{n=0}^{N-1} w^{*}[n]\hat{x}_{\mathrm{BD}}[n-r].
\end{aligned}
\label{eq:rho_expanded_1}
\end{equation}
By expressing these summations as cross-correlation operations, \eqref{eq:rho_expanded_1} simplifies compactly to
\begin{equation}
\begin{aligned}
\rho[r;y,\hat{x}_{\mathrm{BD}}]
&= \sum_{i=0}^{L_{\mathrm{d}}-1} g_{\mathrm{d},i}^{*} R_{x,\hat{x}_{\mathrm{BD}}}(r-\ell_{\mathrm{d},i}) \\
&\quad + \beta \sum_{i=0}^{L_{\mathrm{b}}-1} g_{\mathrm{b},i}^{*} R_{x_{\mathrm{BD}}^{(b)},\hat{x}_{\mathrm{BD}}}(r-\ell_{\mathrm{b},i}) + \zeta_{\rho}(r),
\end{aligned}
\label{eq:rho_expanded_2}
\end{equation}

where $\zeta_{\rho}(r)$ denotes the correlation of \ac{AWGN} onto  $\hat{x}_{\mathrm{BD}}[n]$. From \eqref{eq:rho_expanded_2}, it can be seen that the statistic consists of two terms; $R_{x,\hat{x}_{\mathrm{BD}}}(\cdot)$, due to the \ac{DL}. As shown in Section~\ref{sec:quasi_orthogonality}, this term is weak except near $r=\alpha$ and $r=N-\alpha$ ; hence, a \ac{DL} path with delay $\ell_{d,i}$ generates artifacts around $r=\ell_{d,i}\pm\alpha$. The second term, $R_{x_{\mathrm{BD}}^{(b)},\hat{x}_{\mathrm{BD}}}(\cdot)$, depends on the \ac{BD} bit. For $b=0$, $x_{\mathrm{BD}}^{(0)}[n]=x[n]$, so it reduces to $R_{x,\hat{x}_{\mathrm{BD}}}(\cdot)$ and produces peaks around $r=\ell_{b,i}\pm\alpha$. For $b=1$, $x_{\mathrm{BD}}^{(1)}[n]=x_{\mathrm{BD}}[n]$, and the term becomes $R_{x_{\mathrm{BD}},\hat{x}_{\mathrm{BD}}}(\cdot)$, yielding an autocorrelation-like peak at the effective \ac{BL} delay $r=\ell_{b,i}$, as illustrated in Fig.~\ref{fig:corr1}.

Thus, detection relies on separating the cross-correlation artifacts $R_{x,\hat{x}_{\mathrm{BD}}}(\cdot)$ from the autocorrelation-like peak $R_{x_{\mathrm{BD}},\hat{x}_{\mathrm{BD}}}(\cdot)$. Accordingly, the detection window is chosen within the expected backscatter-delay region while avoiding the \ac{DL}-induced peaks at $r=\ell_{d,i}\pm\alpha$. Selecting $\alpha$ larger than the effective \ac{DL} delay spread provides a wider robust detection window.

\subsection{ Threshold Design for Bit Detection}
This subsection derives the detection threshold by characterizing the correlation
statistic under the null hypothesis. When $b=0$, we define
$\rho_0(r)=\rho[r;y^{(0)},\hat{x}_{\mathrm{BD}}]$.
Since the detection region $\mathcal{R}_b$ excludes the deterministic correlation peaks identified in Section~\ref{sec:quasi_orthogonality}, its residual mean is approximately zero, i.e.,
$\mu_0(r)=\mathbb{E}\{\rho_0(r)\}\approx0$,
$r\in\mathcal{R}_b$.
 Therefore, the variance is
\begin{equation}
\sigma_{0}^{2}(r)
=
\mathbb{E}\left\{
\left|\rho_{0}(r)-\mu_{0}(r)\right|^{2}
\right\}.
\label{eq:rho_h0_var_def}
\end{equation}
Assuming independent channel taps 
$\mathbb{E}\{|R_{x,\hat{x}_{\mathrm{BD}}}(r)|^{2}\}\approx E_x^2/N$,
where $E_x=\frac{1}{N}\sum_{n=0}^{N-1}|x[n]|^{2}$, we obtain
\begin{equation}\small
\sigma_{0}^{2}(r)\approx\frac{E_x^2}{N}
\left(\sum_{i=0}^{L_{\mathrm{d}}-1}|g_{\mathrm{d},i}|^{2}
+|\beta|^{2}\sum_{i=0}^{L_{\mathrm{b}}-1}|g_{\mathrm{b},i}|^{2}\right)+\frac{\sigma_w^{2}}{N^{2}}
\sum_{n=0}^{N-1}|\hat{x}_{\mathrm{BD}}[n-r]|^{2}.
\label{eq:rho_h0_var}
\end{equation}
this reduces to
\begin{equation}
\sigma_{0}^{2}(r)
\approx\frac{E_x^2}{N}\left(
\sum_{i=0}^{L_{\mathrm{d}}-1}|g_{\mathrm{d},i}|^{2}+
|\beta|^{2}\sum_{i=0}^{L_{\mathrm{b}}-1}|g_{\mathrm{b},i}|^{2}+\sigma_w^{2}\right).
\label{eq:rho_h0_var_simple}
\end{equation}
The correlation statistic $\rho_0(r)$ aggregates $N$ finite variance correlation products, while each OFDM sample itself results from the superposition of independent subcarriers. Hence, for sufficiently large $N$, its real and imaginary components are well approximated as Gaussian by the \ac{CLT}. Within $\mathcal{R}_b$,
where $\mu_0(r)\approx0$, we therefore approximate
$\rho_0(r)\sim\mathcal{CN}(0,\sigma_0^2(r))$. Accordingly,
$A(r)=|\rho_0(r)|$ follows approximately a Rayleigh distribution,
whose tail probability is
\begin{equation}
\Pr\left(A(r)>\eta\mid b=0\right)
=\exp\left(-\frac{\eta^{2}}{\sigma_{0}^{2}(r)}\right).
\end{equation}
Thus, for a target false-alarm probability $P_{\mathrm{FA}}$, over the $\mathcal{R}_{\mathrm{b}}$ a threshold is
\begin{equation}
\eta=\sqrt{-\sigma_{0}^{2}\ln\left[1-\left(1-P_{\mathrm{FA}}\right)^{1/|\mathcal{R}_{\mathrm{b}}|}\right]
}.
\label{eq:threshold_window}
\end{equation}
Hence,
\begin{equation}
T=\max_{r\in\mathcal{R}_{\mathrm b}}
|\rho[r;y,\hat{x}_{\mathrm{BD}}]|,\quad
\hat{b}=\mathbf{1}\{T>\eta\}.
\label{eq:decision_rule}
\end{equation}
The overall receiver-side correlation procedure for \ac{BD} signal recovery is summarized in Algorithm~\ref{alg:shufscatter}.
\begin{algorithm}[!t]
\caption{Shuf-Scatter receiver-side Detection}
\label{alg:shufscatter}
\begin{algorithmic}[1]
\State \textbf{Input:} $\hat{y}[n]\leftarrow \eqref{eq:before}, \alpha, \eta$
\State Estimate and equalize the direct-link \ac{OFDM} signal $\hat{y}[n]$ according to \eqref{eq:xhat_time_noise}
\State Recover $\hat{x}[n]$ according to \eqref{eq:xhat_time_noise}
\State Generate the local shuffled reference using \eqref{eq:shuf_scatter_ref}
\State Compute the correlation $\rho[r;y,\hat{x}*{\mathrm{BD}}]$ according to \eqref{eq:rho_expanded_2}
\State Search for peaks within the valid detection window $\mathcal{R}_{\mathrm{b}}$ for $~\alpha< r< N-\alpha$
\State $\hat{b}\gets
\begin{cases}
1, & T>\eta,\\
0, & \text{otherwise},
\end{cases}$
according to \eqref{eq:threshold_window}, \eqref{eq:decision_rule}
\State \textbf{Output:} $\hat{b}$.
\end{algorithmic}
\end{algorithm}

\section{RESULTS AND DISCUSSION}
This section provides a numerical evaluation of the proposed Shuf-Scatter Scheme, focusing on \ac{BER} performance in function of different signal design parameters to validate the theoretical results. Unless otherwise stated, the main simulation parameters are $N=1024$ subcarriers and a \ac{CP} of length $N_{\mathrm{CP}}=64$ samples. The \ac{BD} switching interval is set to $\alpha=32$ samples. The obtained results are benchmarked against the baseline schemes in~\cite{yang2017modulation,devineni2021noncoherent,Qian2016Noncoherent}. 

\begin{figure*}[!t]
\centering
\includegraphics[width=0.85\textwidth, trim=1cm 10cm 1cm 10.25cm,
    clip]{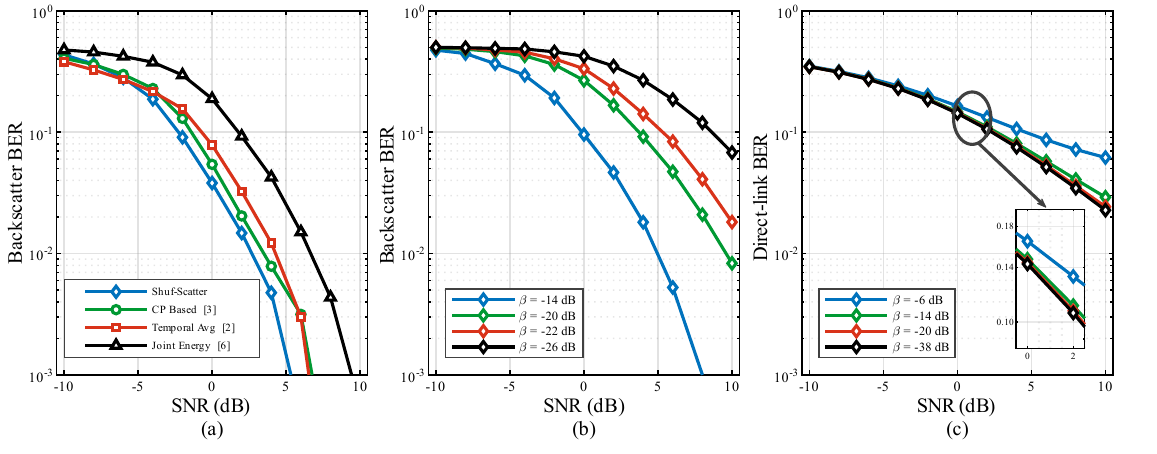}

\caption{\ac{BER} performance of the proposed Shuf-Scatter receiver:
(a) comparison with the benchmark schemes;
(b) impact of the direct-to-backscatter power difference;
(c) recovered direct-link \ac{BER}.}
\label{fig:ber_three_cases}
\end{figure*}

As shown in Fig.~\ref{fig:ber_three_cases}(a), Shuf-Scatter exhibits a steeper \ac{BER} reduction with increasing \ac{SNR} than the considered benchmark schemes. Around $5$~dB, the proposed method reaches a \ac{BER} on the order of $10^{-3}$, while the benchmark schemes remain at higher error levels. This gain results from the deterministic reshuffling of the reflected \ac{OFDM} waveform, which creates a structured correlation signature that can be matched at the receiver. In contrast, the benchmark methods rely on CP-based structure, temporal averaging, or joint-energy observations for backscatter detection. These schemes are more sensitive to the dominant \ac{DLI} because their decision is formed from relatively weak structural or power variations, which can be partially masked by the \ac{DL} component and multipath fluctuations.

Figure.~\ref{fig:ber_three_cases}(b) illustrates the impact of the BD reflection coefficient $\beta$. Varying $\beta$ controls the \ac{DL} to \ac{BL} power imbalance by scaling the strength of the \ac{BL}. Since the reflected power satisfies $P_{\mathrm{BL}}\propto|\beta|^2$, decreasing $|\beta|$ weakens the backscatter link and shifts the \ac{BER} curves toward higher \ac{SNR}. For instance, at $10$~dB, the \ac{BER} is approximately in the $10^{-3}$ range for $\beta=-14$~dB, whereas it remains around $10^{-2}$ for $\beta=-26$~dB. Thus, increasing the \ac{DL} to \ac{BL} power imbalance mainly introduces an \ac{SNR} penalty while preserving the overall \ac{BER} decay trend.

Figure.~\ref{fig:ber_three_cases}(c) reports the \ac{BER} of the recovered \ac{DL} signal for different values of $\beta$. The \ac{DL} \ac{BER} decreases with increasing \ac{SNR} for all considered cases and remains on the order of $10^{-2}$ at $10$~dB. Weaker backscatter reflection generally improves \ac{DL} recovery at high \ac{SNR}, since the backscattered component introduces less interference to the dominant \ac{OFDM} signal.

\section{CONCLUSION AND FUTURE WORK}

This letter proposed Shuf-Scatter, an \ac{OFDM}-compatible ambient backscatter modulation scheme based on controlled time-domain waveform reshuffling. By imposing a deterministic correlation-domain signature on the reflected waveform, Shuf-Scatter reduces the coupling between the weak backscatter component and the dominant \ac{DL} signal.A normalized-correlation receiver was developed to detect the backscatter signal without requiring instantaneous cascaded backscatter-channel \ac{OFDM} or explicit \ac{DL} interference cancellation. Simulations showed that the proposed scheme achieves improved \ac{BER} performance compared with conventional ambient backscatter detectors. Future work will consider adaptive selection of the shuffling interval and multi-device scenario using distinct reshuffling patterns.

\bibliographystyle{IEEEtran}
\bibliography{mybib}
\vfill
\end{document}